\documentclass[aps,pra,preprint,groupedaddress]{revtex4}

\usepackage{amsfonts}
\usepackage{graphicx}
\begin{document}


\title{Quantum adiabatic algorithm for Hilbert's tenth problem:  \\
I.  The algorithm}


\author{Tien D Kieu}
\email[]{kieu@swin.edu.au}
\affiliation{Centre for Atom Optics and Ultrafast Spectroscopy,
Swinburne University of Technology, Hawthorn 3122, Australia}


\date{\today}

\begin{abstract}
We review the proposal of a quantum algorithm for Hilbert's tenth
problem and provide further arguments
towards the proof that: (i) the algorithm terminates after a finite time
for any input of Diophantine equation; (ii) the final ground state which 
contains the answer for the Diophantine equation can be identified as the 
component state having better-than-even probability to be found by measurement
at the end time--even though probability for the final
ground state in a quantum adiabatic process need not monotonically increase
towards one in general.  Presented finally are the reasons why our algorithm
is outside the jurisdiction of no-go arguments previously employed to show that 
Hilbert's tenth problem is recursively non-computable.
\end{abstract}

\pacs{}

\maketitle

\section{\label{sec:intro}Introduction}
We have claimed~\cite{kieu-contphys, kieu-intjtheo, kieu-royal, kieu-spie} to 
have
a quantum algorithm for Hilbert's tenth problem~\cite{hilbert10} despite the
fact that the problem has been proved to be recursively noncomputable.  The 
algorithm
makes essential use of the Quantum Adiabatic Theorem (QAT)~\cite{messiah} and 
other
results (to be further substantiated in this paper) in the framework of
Quantum Adiabatic Computation~\cite{qac} in order to
provide a single, universal procedure, physical or otherwise, which consumes
finite amount of resources and terminates in a finite time and which can tell,
in principle, whether any given Diophantine equation~\footnote{A Diophantine 
equation 
is a polynomial equation having integer coefficients and a finite number of 
unknowns.}
has any non-negative integer solution or not--thus solving Hilbert's tenth.

We refer the readers to the above references and elsewhere for the history
of Hilbert's tenth problem and its paramount importance in Mathematics
and Theoretical Computer Science, as well as in Philosophy.  In the next 
Section,
we summarise the algorithm which employs Quantum Adiabatic Processes (QAP).  
Following that,
we provide some arguments asserting why there is no level crossing, as
required by the QAT, in the spectral flow associated with the QAP, except 
possibly 
at the end points.  This will ensure that our algorithm can terminate in a 
finite time
duration
which is dictated by the smallest, non-vanishing energy gap separating the
instantaneous ground state and relevant excited state in the spectral flow.
We next provide the proof for the criterion identifying the final ground state 
of two-state systems, and then argue that this criterion is also generalisable
to systems of infinitely many states of our algorithm.  We also consider the
algorithm in the context of existing no-go arguments for Hilbert's
tenth problem in order to point out that the algorithm is outside the 
jurisdiction of those arguments.  The paper is concluded with some remarks.

\section{\label{sec:QAC}The Quantum Adiabatic Algorithm}
\subsection{\label{subsection:summary}Statement of the algorithm}
We first introduce the occupation-number state $|n\rangle$, $n = 0, 1, 2, 
\ldots$,
and the creation and annihilation operators $a^\dagger$ and $a$ respectively
\begin{eqnarray}
a |0\rangle &=& 0,\nonumber\\
a |n\rangle &=& \sqrt{n}|n-1\rangle,\nonumber\\
a^\dagger |n\rangle &=& \sqrt{n+1}|n+1\rangle,\nonumber\\
(a^\dagger a) |n\rangle &=& n|n\rangle,
\end{eqnarray}
and the coherent state, with complex number $\alpha$,
\begin{eqnarray}
|\alpha\rangle &=& e^{-\frac{|\alpha|^2}{2}}\sum_{n=0}^{\infty}
\frac{\alpha^n}{\sqrt{n!}}|n\rangle.
\label{coherent}
\end{eqnarray}

Now, given a Diophantine equation with $K$ unknowns, 
\begin{eqnarray}
D(x_1,\ldots,x_K)=0,
\label{equation}
\end{eqnarray}
we provide the following quantum algorithm to decide whether this equation
has any non-negative integer solution or not:
\begin{enumerate}
\item Construct/simulate a physical process in which a system initially
starts with a direct product of $K$ coherent states
\begin{eqnarray}
|\psi(0)\rangle &=& |\{\alpha\}_I\rangle \equiv \bigotimes_{i=1}^K
|\alpha_i\rangle,
\label{initial state}
\end{eqnarray}
and in which the system is subject to a time-dependent Hamiltonian over the
time interval $[0,T]$, for {\em some} time $T$,
\begin{eqnarray}
{\cal H}(t) &\equiv& \left(1 - \frac{t}{T}\right)H_I + \frac{t}{T}H_P,
\label{thehamiltonian}
\end{eqnarray}
with the initial Hamiltonian
\begin{eqnarray}
H_I &=& \sum_{i=1}^K \left(a_i^\dagger - \alpha^*_i\right)
\left(a_i - \alpha_i\right),
\label{initial}
\end{eqnarray}
and the final Hamiltonian
\begin{eqnarray}
H_P &=& \left(D(a^\dagger_1 a_1, \ldots, a^\dagger_K a_K)\right)^2.
\label{final}
\end{eqnarray}
\item Measure/calculate (through the Schr\"odinger equation with the 
time-dependent
Hamiltonian above) the maximum probability to find the system in a
particular occupation-number state at the chosen time $T$,
\begin{eqnarray}
P(T) &=& \max_{|\{n\}\rangle}|\langle\psi(T)|\{n\}\rangle|^2,\nonumber\\
&=&\left|\langle\psi(T)|\{n\}_0\rangle\right|^2,
\label{probability}
\end{eqnarray}
where $|\{n\}_0\rangle$ (which is a direct product of $K$ particular
occupation-number states, 
$\bigotimes_{i=1}^K|n_i^0\rangle$) provides that maximal
probability among all other direct products of $K$ occupation-number states.
\item If $P(T)\le 1/2$, increase $T$ and repeat all the steps above.
\item If $P(T)>1/2$, then $|\{n\}_0\rangle$ is the ground state of $H_P$ 
(assuming
no degeneracy--we discuss the case of degeneracy below) and we can terminate the
algorithm and deduce a conclusion from the fact that:
\begin{eqnarray}
H_P|\{n\}_0\rangle = 0 {\mbox{ {\bf iff} the equation~(\ref{equation}) has a 
non-negative integer solution.}}
\end{eqnarray}
\end{enumerate}

We refer to~\cite{kieu-contphys, kieu-intjtheo, kieu-royal} for the motivations 
and
discussions leading to the algorithm above.  Note
that it is crucial that $H_I$ does not commute with $H_P$,
\begin{eqnarray}
[H_I,H_P] &\not =& 0,
\label{commutation}
\end{eqnarray}
and of course both Hamiltonians are of infinite dimensions.  (This
condition can be relaxed to that of unbounded dimensions, see later.)

To remove any possible degeneracy of the ground state for any $H_P$, we can
always
introduce to $H_P$
a symmetry-breaking term of the form $(\gamma a^\dagger_i+\gamma^* a_i)$
for some $i$ in the limit ${|\gamma|}\to 0$~\cite{kieu-royal}.  This term
destroys
the symmetry generated from the commutation between $H_P$ and the 
occupation-number
operators.  However, we can always recover the symmetry in the limit and modify
the algorithm above slightly to reach an answer for
the Diophantine equation in question.
Note also that the arguments
of the Section~\ref{sec:nondegenerate} below can be easily modified to
accommodate this
symmetry-breaking term to ensure there is no ground-state degeneracy in $0\le
t/T <1$.

\subsection{\label{subsec:probabilistic}Its probabilistic nature}
We will prove in the next Section that there exists a finite $T$ whence the
algorithm can be terminated.
But it is important to recognise hereby that our algorithm is probabilistic in
nature.
As the halting criterion, the probability $P(T)$ of~(\ref{probability})
for certain state need to be more than one-half.
If the algorithm is implemented by certain physical process then
we can approximate this probability by the relative
frequency that a particular state is obtained in repeating the measurements
many times over.  The Weak Law of Large Numbers, however, can only assert
that this frequency is within a distance $\epsilon$ of $P(T)$ {\em with certain
probability} of at least $1-\delta$.
It is this probability $1-\delta$ that gives the quantum algorithm its
probabilistic nature.
Both $\epsilon$ and $\delta$ are dependent on the number of measurement
repetitions from which the frequency is obtained--see, for
example,~\cite{probtheory} for a statement and proof of the theorem.
In general, we can always reduce $\epsilon$ and $\delta$ to arbitrarily
close to zero by increasing the number of repetitions $L$ appropriately,
\begin{eqnarray}
L > 1/(4\epsilon^2\delta).
\label{L}
\end{eqnarray}

As an anticipation of the numerical simulations of the algorithm on recursive
computers~\cite{kieu-spie,kieu-numer}, we mention now that the probabilistic
nature
of the algorithm is manifest differently there through the extrapolation to zero
step size in solving the Schr\"odinger equation.  Such extrapolation is
necessary for a correct truncation, for a given Diophantine equation,
of the dimensions of the Hilbert space involved.

We will later come back to this probabilistic nature in
Section~\ref{sec:non-recursive}
in the discussion of no-go arguments for Hilbert's tenth problem.   In ending
this Section,
we wish to point out here that, contrary to common misunderstanding,
probabilistic
computation in general is {\em not} equivalent in terms of computability to
Turing
computation~\cite{OrdKieu-prob}.

\section{\label{sec:nondegenerate}No level crossing for the ground state}
One of the sufficient conditions for the Quantum Adiabatic
Theorem~\cite{messiah}
is that the relevant level, which in our case is the instantaneous
ground state, doesnot cross with any other level as time progresses.
In~\cite{kieu-royal}
we have given some arguments for no level crossing for the time-dependent
Hamiltonian
\begin{eqnarray}
{\cal H}(sT) &=& (1-s)H_I + sH_P,
\label{adiahamiltonian}
\end{eqnarray}
with the reduced time $s$ in the interval $s\in(0,1)$.  Now we present below
some
other arguments adapted from those
of Ruskai~\cite{Ruskai:2002}, which in turn are based on Perron-Frobenius's
theorem~\cite{hornj85} for finite but unbounded number of dimensions, 
and on Reed-Simon's theorems~\cite{ReedSimon}
for infinite number of dimensions.

Let us consider the exponential operator $\exp(-{\cal H}(sT))$,
it can be expressed by the Lie-Trotter product formula as
\begin{eqnarray}
{e}^{-{\cal H}(sT)} &=& \lim_{M\to \infty}\left({e}^{-\frac{s}{M}H_P
-\frac{1-s}{M}\sum_i (a^\dagger_i a_i + \alpha^*_i \alpha_i)}
\left[
{e}^{\frac{1-s}{M}\sum_i \alpha_i a^\dagger_i}
{e}^{\frac{1-s}{M}\sum_i \alpha^*_i a_i}
\right]
\right)^M.
\label{approx}
\end{eqnarray}
For general operators in dimensionally infinite Hilbert spaces, the convergence
in the above is understood to be the {\em strong convergence} of the rhs to
the lhs~\footnote{A sequence of operators $A_n$ is said to converge strongly to
$A$ in a Hilbert space $H$ if the norm $\|(A_n-A)|f\rangle\|$ converges to zero
for $n\to\infty$ for any $|f\rangle \in H$.}.

\subsection{Finite but unbounded number of dimensions}
Our algorithm of the last Section actually requires only sufficiently large but
finite Hilbert space with a truncated basis consisting of occupation-number
vectors, $\{|n\rangle: n=0,1,\ldots, N\}$, for some $N$ beyond which higher
occupation-number states contribute negligibly to the dynamics of low-lying 
states relevant to our problem.  This follows from the facts that the normalised 
wavefunction at any given time has a support which
spreads significantly only over a finite range of occupation-number states, and
that the explicit coupling, and thus the influence, between one
particular instantaneous eigenstate and other diminishes significantly outside
some finite range of occupation-number states (as can be seen
through a set of differential equations in~\cite{kieu-royal}
connecting instantaneous eigenstates of ${\cal H}(t)$ at different time $t$).
We will exploit these facts fully in the numerical simulations of the quantum
algorithm~\cite{kieu-spie,kieu-numer}.
How large is sufficient, and thus how much the truncation $N$ should be, depend
very much on the particular Diophantine equation being investigated.  
Such truncation will be discussed and implemented in the numerical ssimulations
of the quantum algorithm~\cite{kieu-spie,kieu-numer}.
In the last Section, we need not worry about the size of $N$ as
we have employed dimensionally infinite Hilbert space.
However, for any arbitrary and unbounded $N$, the
Hamiltonian~(\ref{adiahamiltonian})
above has a representation which is a finite square matrix and in which all the
off-diagonal elements are contained in $H_I$.

The elements of the matrix product in the square brackets on the rhs
of~(\ref{approx}), can be expressed in the truncation to an {\em arbitrary}
$N$ (where $a^\dagger |N\rangle = 0$) as
\begin{eqnarray}
\langle m|
{e}^{\beta_i a^\dagger_i}
{e}^{\beta^*_i a_i}
|n\rangle
&=&
\sum_{q=0}^N
\langle m|
{e}^{\beta_i a^\dagger_i}
|q\rangle\langle q|
{e}^{\beta^*_i a_i}
|n\rangle,\nonumber\\
&=&
\sum_{q=0}^N
\langle m|
\sum_{k=0}^\infty \frac{(\beta_i a^\dagger_i)^k}{k!}
|q\rangle\langle q|
\sum_{l=0}^\infty \frac{(\beta_i^* a_i)^l}{l!}
|n\rangle,
\end{eqnarray}
where 
\begin{eqnarray}
\beta_i= \frac{1-s}{M}\alpha_i.
\end{eqnarray}
It then follows that for $\{\alpha_i\}$ real and positive, and for $0\le s<1$,
the lhs above is always non-zero and positive.  For example, when $N>m>n$, 
at least the term with $k=m-n$ and $l=0$ will contribute
nonvanishingly and positively to the sum of non-negative terms on the rhs to
ensure the matrix element on the lhs is always positive.

In the basis of occupation-number states,
the first factor in the bracket of the rhs of~(\ref{approx}) is a diagonal matrix,
whose net effect is to multiply each row of the matrix product in square 
brackets in~(\ref{approx})
by non-zero, positive numbers (being exponentials).  As a result, for
$\alpha_i = \alpha_i^* > 0, {\mbox{ for }} i = 1, 2, \ldots, K$,
the product in the bracket on the rhs
of~(\ref{approx}) has only positive elements, and so has
its $m$-th power.
Thus, Perron-Frobenius's theorem~\cite{hornj85} for (finite) matrices
with strictly positive elements can be applied to
the lhs of~(\ref{approx}) to confirm that the largest eigenvector
of ${\rm e}^{-{\cal H}(sT)}$ is unique for these values of
$\{\alpha_i\}$.  That is, ${\cal H}(sT)$ itself
has a unique ground state for $0\le s < 1$ (since, by
construction, ${\cal H}(0) = H_I$ has a nondegenerate ground state).

At $s=1$, the matrix ${\rm e}^{-\frac{1-s}{M}H_I}$ becomes the identity matrix
which has zero non-diagonal elements and becomes reducible, violating the
conditions of Perron-Frobenius's theorem and thus allowing the possibility that
${\cal H}(T) = H_P$ has a degenerate ground state.

\subsection{Infinite number of dimensions}
To handle the infinite dimensions of the operator product in
the square brackets of~(\ref{approx}) directly, that is, without any truncation,
we can employ the so-called holomorphic representation~\cite{faddeev,qft} to
arrive at
\begin{eqnarray}
\langle m|
{e}^{\beta_i a^\dagger_i}
{e}^{\beta^*_i a_i}
|n\rangle &=&
\int\frac{d\bar\eta d\eta}{2\pi i}
\frac{d\bar\omega d\omega}{2\pi i}
\frac{d\bar\xi d\xi}{2\pi i}
\frac{\eta^m}{\sqrt{m!}}
e^{-\bar\eta\eta}
e^{\bar\eta\omega + \bar\eta\beta_i}
e^{-\bar\omega\omega}
e^{\bar\omega\xi + \beta_i^*\xi}
e^{-\bar\xi\xi}
\frac{\bar\xi^n}{\sqrt{n!}},
\nonumber\\
&=&\frac{1}{\sqrt{m!n!}}
\left(\frac{\partial}{\partial\bar s}\right)^m
\left(\frac{\partial}{\partial s}\right)^n
\int\frac{d\bar\eta d\eta}{2\pi i}
\frac{d\bar\xi d\xi}{2\pi i}
\left. e^{-\bar\eta\eta
+ \bar\eta\xi + \bar\eta\beta_i
+ \beta_i^*\xi
-\bar\xi\xi
+ \bar s\eta +\bar\xi s}
\right|_{s=\bar s=0},\nonumber\\
&=&\frac{1}{\sqrt{m!n!}}
\left.
\left(\frac{\partial}{\partial\bar s}\right)^m
\left(\frac{\partial}{\partial s}\right)^n
\exp\{\bar s\beta_i + \bar s s + \beta_i^* s\}
\right|_{s=\bar s=0},
\label{holomorphic}
\end{eqnarray}
where we have integrated over $\{\bar\omega,\omega\}$ and introduced the source
terms $\bar s$ and $s$ in the second line, and done a Gaussian intergration to 
get the last line.  It can be seen from~(\ref{holomorphic}),
that the matrix element always is strictly positive for any positive 
integers $m$ and $n$ provided that $0\le s<1$ and $\alpha_i$ is real and positive.
From this we are led to the conclusion that 
\begin{eqnarray}
\langle n|{e}^{-\frac{s}{M}H_P
-\frac{1-s}{M}\sum_i (a^\dagger_i a_i + \alpha^*_i \alpha_i)}
\left[
{e}^{\frac{1-s}{M}\sum_i \alpha_i a^\dagger_i}
{e}^{\frac{1-s}{M}\sum_i \alpha^*_i a_i}
\right]|m\rangle =
\nonumber\\
{e}^{-\frac{s}{M}D^2
-\frac{1-s}{M}\sum_i (n_i + |\alpha_i|^2)}
\langle n|
{e}^{\frac{1-s}{M}\sum_i \alpha_i a^\dagger_i}
{e}^{\frac{1-s}{M}\sum_i \alpha^*_i a_i}|m\rangle > 0,
\label{positive2}
\end{eqnarray}
and consequently that the lhs of~(\ref{approx}) has positive matrix elements
for real and positive $\alpha_i$.

For the case of {\em bounded} operator on the lhs of~(\ref{approx}), we can adopt 
a theorem of Reed and
Simon~\footnote{Theorem XIII.44 in~\cite{ReedSimon}, pp. 204--205.},
a cut-down version of which can be rephrased for our purpose as
\begin{eqnarray}
{\mbox{${\cal H}$ has a nondegenerate ground state}}
{\mbox{ \bf iff }}
{\mbox{$e^{-a{\cal H}}$ is {\em positive improving} for all $a>0$.}}
\nonumber
\end{eqnarray}
In our occupation-number basis, having $e^{-a{\cal H}}$ 
{\em positive improving} for all $a>0$ is equivalent to having matrix element
$\langle m|e^{-a{\cal H}}|n\rangle$ strictly positive for all $a>0$ and
for all positive integers $m$, $n$.  But the nonvanishing of such matrix elements
follows simply and directly from the result in~(\ref{positive2}) and the
strong convergence in~(\ref{approx}). Thus, finally,
the non-degeneracy of the ground state of the dimensionally infinite
operator ${\cal H}(sT)$ also follows.

We can in fact prove a stronger result, by
invoking another theorem of Reed and Simon~\footnote{Theorem XIII.43 
in~\cite{ReedSimon}, pp. 202--204.}, that {\em all} the eigenvectors of ${\cal H}(sT)$,
not just the ground state,
are non-degenerate for real and positive $\alpha_i$'s and for $0\le s < 1$.

\subsection{Extension to $\alpha_i\in\mathbb C$}
Having established the above results for real and positive $\alpha_i$'s, we now 
extend them to all complex-valued but nonvanishing $\alpha_i$'s.
Indeed, having all $\alpha_i$ vanishing would make $H_I$ commute with $H_P$, leading
in general to an unwanted crossing of the ground state of ${\cal H}(t)$ at some 
$t\in(0,T)$.

The crucial point is to note that the Hamiltonian~(\ref{thehamiltonian}) is
invariant under the following phase transformations,
for real and arbitrary $\theta_i$,
\begin{eqnarray}
\left\{
\begin{array}{lll}
\alpha_i &\to& e^{i\theta_i}\alpha_i,\\
a_i &\to& e^{i\theta_i}a_i.
\end{array}
\right.
\label{phasetrans}
\end{eqnarray}
Denoting
\begin{eqnarray}
\bar a_i &\equiv& e^{i\theta_i}a_i = \left(e^{-i\theta_i a^\dagger_i a_i}\right)
a_i\left(e^{i\theta_i a^\dagger_i a_i}\right),
\end{eqnarray}
we see that these operators also satisfy the canonical commutation relation,
$[\bar a_i,{\bar a}^\dagger_j]=\delta_{ij}$, and we can thus construct a 
unitarily related Fock space, with the basis consisting of $|\{\bar n_i\}\rangle
=\bigotimes_{i=1}^K|\bar n_i\rangle$, such that
\begin{eqnarray}
|\bar n_i\rangle &=& e^{-i n_i \theta_i}|n_i\rangle,\\
({\bar a}^\dagger_i \bar a_i) |\bar n_i\rangle &=& n_i|\bar n_i\rangle.\nonumber
\end{eqnarray}

Now, with nonvanishing complex-valued $\alpha_i = e^{-i\theta_i}|\alpha_i|$, we
can perform the phase transformation~(\ref{phasetrans}) to obtain an
equivalent  Hamiltonian
containing only real and positive $|\alpha_i|$, and $\bar a_i$.  The arguments of the
two subsections above, for both finite and infinite number of dimensions, will 
carry through but this time with the unitarily transformed occupation-number states 
$|\{\bar n_i\}\rangle$.  Once again, the nondegeneracy of the ground state
of ${\cal H}(t)$ is assured, but this time extended for nonvanishing complex-valued
$\alpha_i$'s.

All of the above agrees with our alternative arguments in~\cite{kieu-royal} which lead
to the nondegeneracy of the ground state of ${\cal H}(sT)$ for $0\le s <1$.  The
possible degeneracy of $H_P$ can then be handled as discussed in the last
Section.  It follows next from the Quantum Adiabatic Theorem that if the 
instantaneous ground
state never crosses with any other state for the whole of the spectral flow then
it takes a non-vanishing rate of change $1/T$, and thus only a {\em finite} time
$T$, for the probability of the final ground state to evolve arbitrarily closed
to one.  This implies that our quantum algorithm can always be terminated in a finite
time.  The next Section gives the criterion for finding that terminating time.

\section{\label{sec:identification}Identifying the groundstates}
\subsection{\label{subsec:overall}The overall picture}
The crucial step of any quantum adiabatic algorithm is the identification of the
ground state
of the final Hamiltonian, $H_P$.  Normally it is identified as the
probabilistically
dominant state obtained for an adiabatic evolution time $T$ sufficiently long as
asserted
by the Quantum Adiabatic Theorem~\cite{messiah}.
In our case we do not in advance know in general how long is sufficiently long
(the Theorem offers no direct help here);
all we can confidently know is that for each
Diophantine equation and each suitable $\{\alpha_i\}$ there is a {\em finite}
evolution time
(the finiteness is due to the non-degeneracy of
the instantaneous ground state--see last Section) after which the adiabatic
condition
is satisfied.  We thus have to find another criterion to identify the ground
state.

We will make full use of the fact that the eigenstates of $H_P$ are by
construction just the occupation-number states,
$\{|\{n\}\rangle=\bigotimes_{i=1}^K|n_i\rangle: (a^\dagger_i a_i|n_i\rangle= n_i
|n_i\rangle) \& (n_i=0,1,\ldots)\}$, among which is the final ground state to
be identified.

The identification criterion we have found can be stated as:
\begin{quotation}
{\it The ground state of $H_P$ is the component
state $|\{n\}_0\rangle$ whose measurement
probability is more than $\frac{1}{2}$ after the evolution for
some time $T$ of the initial ground state $|\{\alpha\}_I\rangle$ according to
the Hamiltonian~(\ref{adiahamiltonian})},
\end{quotation}
\begin{eqnarray}
|\langle\psi(T)|\{n\}_0\rangle|^2 > \frac{1}{2}, {\mbox{ for some }} T,
{\mbox{ \bf if and only if }} |\{n\}_0\rangle
{\mbox{ is the ground state of }} H_P.
\end{eqnarray}
We can recognise the Quantum Adiabatic Theorem in the {\em `if'} part
of the statement above; and we need only prove the {\em `only if'} part.
In general, it suffices that the initial ground state
$|g_I\rangle$ of some $H_I$ should not have any dominant component in the
occupation-number eigenstates $|\{n\}\rangle$ of $H_P$,
\begin{eqnarray}
|\langle g_I|\{n\}\rangle|^2 \le \frac{1}{2},\forall \{n\}.
\label{nodominant}
\end{eqnarray}
Our choice of initial coherent state $|\{\alpha\}_I\rangle$~(\ref{initial
state})
clearly satisfies this condition.

Consequently, we only need to increase the
evolution time $T$ until one of the occupation-number states is obtained
at time $T$ with a probability of more than
$\frac{1}{2}$ then this will be our much desired ground state.

We will prove this criterion for two-state systems in the next Subsection
and then argue that it is also applicable
for systems of finitely many and of infinite number of states because
among those states only two states, which may or may not be the instantaneous
ground state and first
excited state, become dominantly relevant at any instant of time-- provided we
start
out with the ground state of the initial Hamiltonian $H_I$.

\subsection{\label{subsec:two}Two-state systems}
We divide the proof for two-state systems in turn into three parts:
\begin{itemize}
\item In the first one, we
show (Eq.~(\ref{result}) below) that the maximum probability for the state which
is {\em not}
the final ground state {\em cannot} be more than $\frac{1}{2}$ for any evolution 
time
$T$--subject to certain condition stated below (and which is also satisfied by 
our 
choice of $H_I$ in the generalisation to dimensionally infinite Hilbert spaces).  

Note that the probability need not be monotonic function of $T$.  
\item In the second part, we appeal to the
Quantum Adiabatic Theorem~\cite{messiah} which asserts that eventually for 
sufficiently
large $T$ (i.e. for sufficiently slow evolution rate $1/T$) the probability for
the final ground state approaches one.  
\item Then by combining the two parts above, we can,
without the need of knowing {\em a priori} how slow is sufficiently slow for
the evolution rate $1/T$,
conclude that eventually the probability of one state will rise above 
$\frac{1}{2}$ 
as $T$ increases and that state must be the ground state.  
\end{itemize}

It suffices to establish the first part of the arguments above.
Let $|g(t)\rangle$ and $|e(t)\rangle$
be the instantaneous eigenstates of ${\cal H}^{(2)}(t)$, the two-dimensional
counterpart 
of~(\ref{thehamiltonian}).  They are
related to the eigenstates of ${\cal H}^{(2)}(t +\delta t)$ by 
\begin{eqnarray}
|g(t)\rangle &=& \cos\left(\beta(t)/2\right)|g(t+\delta t)\rangle +
\sin\left(\beta(t)/2\right)
|e(t+\delta t)\rangle,\nonumber\\
|e(t)\rangle &=& -\sin\left(\beta(t)/2\right)|g(t+\delta t)\rangle + 
\cos\left(\beta(t)/2\right)
|e(t+\delta t)\rangle,
\label{differentt}
\end{eqnarray}
where
\begin{eqnarray}
\tan(\beta(t)) 
&=& \frac{2\langle e(t)|\partial_t {\cal H}^{(2)}(t)|g(t)\rangle}
{\langle e(t)|{\cal H}^{(2)}(t)|e(t)\rangle
-\langle g(t)|{\cal H}^{(2)}(t)|g(t) \rangle}\delta t + O(\delta t^2)
\label{tan}
\end{eqnarray}
by recalling that ${\cal H}^{(2)}(t+\delta t)
=  {\cal H}^{(2)}(t) + \delta t \partial_t {\cal H}^{(2)} + O(\delta t^2)$,
where
\begin{eqnarray}
{\cal H}^{(2)}(t) &=& \left(1-\frac{t}{T}\right)H_I^{(2)} + 
\frac{t}{T}H_P^{(2)},\nonumber\\
H_I^{(2)} &=& \epsilon_g|g(0)\rangle\langle g(0)|
+ \epsilon_e|e(0)\rangle\langle e(0)|, 
\label{twohamiltonian} \\
H_P^{(2)} &=& \Upsilon_g|g(T)\rangle\langle g(T)|
+ \Upsilon_e|e(T)\rangle\langle e(T)|, \nonumber
\end{eqnarray}
for some $\epsilon$'s and $\Upsilon$'s and such that~(\ref{commutation}) is 
observed,
\begin{eqnarray}
[H_I^{(2)},H_P^{(2)}] &\not =& 0.
\label{nocommutation}
\end{eqnarray}
Note that by redefining the phases
of $|e(t)\rangle$ and $|g(t)\rangle$ in~(\ref{differentt}) we can always choose 
\begin{eqnarray}
0\le\beta(t)/2 <\pi/2.
\label{range}
\end{eqnarray}

Assuming that at time $(t + \delta t)$,
\begin{eqnarray}
|\psi(t + \delta t)\rangle &\sim& \cos(\phi(t))|g(t)\rangle + 
{\rm e}^{-i\theta(t)}\sin(\phi(t))|e(t)\rangle,
\label{timet}
\end{eqnarray}
for some $\theta(t)$ and $\phi(t)$ such that 
\begin{eqnarray}
\tan(\phi(t))\ge 0, 
\label{positive}
\end{eqnarray}
then
\begin{eqnarray}
|\psi(t+2\delta t)\rangle &=& e^{-i{\cal H}(t+\delta t)\delta t}
|\psi(t+\delta t)\rangle,\nonumber\\
&\sim& \left(\cos(\phi(t))\cos(\beta(t)/2)-
{\rm e}^{-i\theta(t)}\sin(\phi(t))\sin(\beta(t)/2)\right)|g(t+\delta t)\rangle
\label{psi} \\
&&+{\rm e}^{-i\theta(t + \delta t)} \left(\cos(\phi(t))\sin(\beta(t)/2)+
{\rm e}^{-i\theta(t)}\sin(\phi(t))\cos(\beta(t)/2)\right)|e(t+\delta t)\rangle,
\nonumber
\end{eqnarray}
by the use of~(\ref{timet}) and~(\ref{differentt}).
According to (\ref{psi}),
the probability to be found in the ground state at time $(t+2\delta t)$ is
\begin{eqnarray}
\cos^2(\phi(t+\delta t)) &\equiv&
\left|\cos(\phi(t))\cos(\beta(t)/2)-
{\rm e}^{-i\theta(t)}\sin(\phi(t))\sin(\beta(t)/2)\right|^2,\nonumber\\
&\ge& \cos^2(\phi(t)+\beta(t)/2);
\end{eqnarray}
where we have made use of (\ref{positive}) to derive the last inequality.
Similarly, the probability to be found in the excited state is
\begin{eqnarray}
\sin^2(\phi(t+\delta t)) &\equiv&
\left|\cos(\phi(t))\sin(\beta(t)/2)+
{\rm e}^{-i\theta(t)}\sin(\phi(t))\cos(\beta(t)/2)\right|^2,\nonumber\\
&\le& \sin^2(\phi(t) + \beta(t)/2),\nonumber\\
&\le& \sin^2\left(\sum_{n=0}^{t/\delta t} \beta(n\delta t)/2\right),
\label{sine}
\end{eqnarray}
where we have used the arguments of induction and the infinitesimality in
$\delta t$ of $\beta(n\delta t)$ to arrive at the last line 
above~\footnote{The induction proof can be constructed from the fact that
$\phi + \beta/2 < \pi/2$, which in turn follows from $\phi<\pi/2$, as in~(\ref{positive}),
and from $\tan(\phi+\beta/2) = \tan(\phi) + \tan(\beta/2) + O(\delta t^2)
> 0$, as asserted by~(\ref{tan}) and~(\ref{range}).}.

Denoting
\begin{eqnarray}
\omega(\tau) \equiv\lim_{\delta t \to 0}
\sum_{n=0}^{\tau/\delta t} \beta(n\delta t),
\end{eqnarray}
we now sum up the infinitesimals $O(\delta t)$ of (\ref{tan})
to have
\begin{eqnarray}
\tan\left(
\omega(\tau)\right)
&=& 
\int_0^\tau \frac{2\langle e(t)|\partial_t {\cal H}^{(2)}(t)|g(t)\rangle}
{\langle e(t)|{\cal H}^{(2)}(t)|e(t)\rangle
-\langle g(t)|{\cal H}^{(2)}(t)|g(t) \rangle}dt.
\label{integral}
\end{eqnarray}

Our next step is to prove that the integrand in~(\ref{integral}) is never 
zero for $t\in(0,T)$.  It is a proof by contradiction if the opposite is 
assumed.
That is, assuming that there exists a time $t_0$ at which the numerator of the 
integrand vanishes
\begin{eqnarray}
0 = \langle e(t_0)|\partial_t {\cal H}^{(2)}(t_0)|g(t_0)\rangle =
\langle e(t_0)|H_P^{(2)}-H_I^{(2)}|g(t_0)\rangle/T, 
\label{numerator}
\end{eqnarray}
by the use of~(\ref{twohamiltonian}).  This means that the combined operator
$H_P^{(2)}-H_I^{(2)}$ is diagonal
in the basis $\{|g(t_0)\rangle, |e(t_0)\rangle\}$.  On the other hand, the 
operator
${\cal H}^{(2)}(t_0) = H_I^{(2)} + (t_0/T)(H_P^{(2)}-H_I^{(2)})$ is 
automatically diagonal 
in the same basis by virtue of its very definition.  Thus, these two operators 
must commute
\begin{eqnarray}
0 &=& [{\cal H}^{(2)}(t_0), H_P^{(2)}-H_I^{(2)}],\nonumber\\
&=& [H_I^{(2)},H_P^{(2)}],
\label{commute}
\end{eqnarray}
which contradicts our assumption concerning $H_I$ and $H_P$
in~(\ref{nocommutation}).
Thus there cannot exist any such $t_0$ that satisfies~(\ref{numerator}).

Having established that the integrand is never zero, we can find out its sign by
evaluating it at $t=0$, say.  The integral~(\ref{integral}) likewise never
vanishes for
any $\tau$ and assumes the same sign as its integrand.

The case of interest is when the sign is positive
\begin{eqnarray}
\tan\left(
\omega(\tau)
\right) \ge 0,\,
{\mbox{\rm  for all }} \tau\in(0,T),
\label{angle}
\end{eqnarray}
which implies that 
\begin{eqnarray}
0\le \omega(\tau)
\le\pi/2.  
\label{firstquad}
\end{eqnarray}
This last line follows from the continuity
in $\tau$ which smoothly connects $\omega(\tau)$ at $\tau\not=0$ to $\omega(0)$:
\begin{itemize}
\item from~(\ref{range}) we know that $\beta(0)$ is in the range $[0,\pi)$;  
\item thus, $\omega(0)$, which equals $\beta(0)$, has to be less than $\pi/2$ 
for~(\ref{angle}) 
to hold at $\tau = 0$;
\item consequently, as $\tan(\omega(\tau))$ never for any $\tau$ is negative,
$\omega(\tau)$ has to be less than $\pi/2$, by the continuity of $\omega(\tau)$ 
as a 
function of $\tau$, for~(\ref{angle}) to hold at all $\tau$.  (This last 
statement can
also be proved by employing the method of proof by contradiction.)
\end{itemize}

Thus, we are led from~(\ref{firstquad}) and~(\ref{sine}) to an upper bound on 
the probability 
of the excited state at all time
\begin{eqnarray}
\sin^2(\phi(\tau)) \le
\sin^2\left(
\omega(\tau)/2
\right) \le \sin^2(\pi/4) = \frac{1}{2}.
\label{result}
\end{eqnarray}
This completes our main arguments for this Section.

Note that the condition~(\ref{angle}) is crucial here.  It is satisfied for
\begin{eqnarray}
|\langle g(0)|e(T)\rangle|^2 \le \frac{1}{2},
\label{condition}
\end{eqnarray}
and the system is initially in the ground state $|\psi(0)\rangle =
|g(0)\rangle$.
If the opposite of the above is true then we can show that there exists
some range of $T$ such that~(\ref{result}) does not hold. In fact, this can
easily 
be seen in the sudden approximation when the rate of change $1/T$ is large
\begin{eqnarray}
\lim_{T\to 0} |\psi(T)\rangle &\approx& |\psi(0)\rangle = |g(0)\rangle,\nonumber 
\end{eqnarray}
from which, assuming the opposite of~(\ref{condition}),
\begin{eqnarray}
|\langle \psi(T)|e(T)\rangle|^2 &\approx& 
|\langle g(0)|e(T)\rangle|^2 > \frac{1}{2},
\end{eqnarray}
which is the opposite of the result~(\ref{result}).  (The existence of a range 
of values
of such $T$ immediately follows from the continuity of $|\psi(T)\rangle$ as a
function of $T$.) By solving the Schr\"odinger equations with time-dependent
Hamiltonians, we illustrate in Figure 1 both the cases 
when~(\ref{result}) is and is not held depending on whether 
the condition~(\ref{condition}) is satisfied or not.
\begin{figure}
\includegraphics{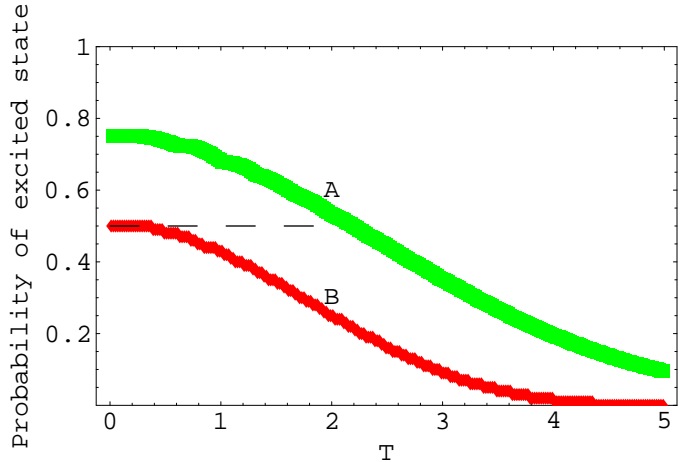}%
\caption{\label{Fig1}Probability distributions as functions
of $T$ for two-state systems, illustrating the cases
of (A) $|\langle g(0)|e(T)\rangle|^2 = 3/4 > 1/2$
and (B) $|\langle g(0)|e(T)\rangle|^2 = 1/2$.
For the first case, 
there is a range of $T$ which violates~(\ref{result}),
while in the latter we always have~(\ref{result}). 
The excited-state probability always approaches zero for sufficiently
large $T$ according to the Quantum Adiabatic Theorem.}
\end{figure}

When the above is generalised to dimensionally infinite systems
of the quantum algorithm for Hilbert's tenth problem,
the counterpart of~(\ref{differentt}) is the set of differential equations
connecting the sets of instantaneous eigenstates of ${\cal H}(t)$ at different
time $t$~\cite{kieu-royal}. There,
our choice of the coherent state~(\ref{initial state}), as the ground state in
which
the system
initially has to be, entails that the condition~(\ref{condition}) is always
satisfied,
since for any $n$ and $\alpha$
\begin{eqnarray}
|\langle \alpha|n\rangle|^2 &=& e^{-|\alpha|^2}\frac{|\alpha|^{2n}}{n!} <
\frac{1}{2}.
\end{eqnarray}
And we expect that, in these infinite systems too, the probability to be
subsequently
found in any particular excited state cannot be greater than $\frac{1}{2}$ at
anytime.  Such arguments have also been numerically confirmed in several
simulations of our algorithm~\cite{kieu-spie, kieu-numer} and also of a modified
version~\cite{columbia}.

\section{\label{sec:non-recursive}How can we compute the non-computable?}
When proposed the tenth problem in 1900, Hilbert himself never 
anticipated the link it would have with what is the Turing halting problem
of the yet-to-be-born field of Theoretical Computer Science.  The Turing halting 
problem was only introduced and solved in 1937 by Turing, and the link of
equivalence between the two problems was only established in 1972~\cite{hilbert10}.

\subsection{\label{Turing}The Turing halting problem}
The question of the Turing halting problem can be phrased as whether there
exists a universal process according to which it can be determined by a finite
number of operations if any given Turing machine would eventually halt (in
finite time) starting  with some specific input.  Turing raised this problem in
parallel similarity to the G\"odel's Incompleteness Theorem and settled it with
the result that there exists no such recursive universal procedure.  The proof
is based on Cantor's diagonal arguments, also employed in the proof of the
Incompleteness Theorem.

The proof is by contradiction starting with the assumption that there exists a
recursive (and hence Turing computable) single-valued halting function $h(p,i)$
which accepts two integer inputs: $p$, the G\"odel encoded integer for the Turing
machine in consideration, and  $i$, the G\"odel encoded integer for the input for $p$,
\begin{eqnarray}
h(p,i)&=&\left\{
\begin{tabular}{l}
0 \mbox{ if $p$ halts on input $i$};\\
1 \mbox{ otherwise}.
\end{tabular}
\right.
\label{Cantor}
\end{eqnarray}
One can then construct a program $Turing(n)$ having one integer argument $n$ in such a
way that it calls the function $h(n,n)$ as a subroutine and then halts if and only
if $h(n,n) = 1$.  In some made-up language:
\begin{eqnarray}
\begin{tabular}{ll}
 &Program $Turing$\\
 &input $n$\\
10& call $h(n,n)$\\
 &if $h(n,n) = 0$ goto 10\\
 &stop\\
 &end
\end{tabular}
\end{eqnarray}

Let $t$ be the G\"odel encoded integer for $Turing$; we now apply the assumed halting
function $h$ to $t$ and $n$, then clearly:
\begin{eqnarray}
h(t,n) = 0 && \mbox{\bf  if and only if }
Turing \mbox{ halts on } n\nonumber\\
&&\mbox{\bf  if and only if } h(n,n) =1,
\label{T1}
\end{eqnarray}
from which a contradiction is clearly manifest once we choose $n = t$.

The elegant proof above was only intended by Turing for the
non-existence of a
recursive halting function.  Unfortunately, some has used this kind of arguments
to argue that there cannot exist any halting function in general!
We have pointed out elsewhere~\cite{OrdKieu-diag} the
fallacies in such use, and considered carefully the implicit assumptions of
Cantor's diagonal arguments.  See also the Subsection~\ref{jurisdiction} below.

\subsection{The equivalence between Hilbert's tenth and the Turing halting
problem}
It is easy to see that if one can
solve the Turing halting problem one can then solve Hilbert's tenth problem.
This is accomplished by constructing a simple program that systematically
searches for the zeros of a given Diophantine equation by going through the
non-negative integers one by one and stops as soon as a solution is found.  The
Turing halting function (existed by assumption) can then be applied to that
program to see if it ever halts or not.  It halts if and only if the Diophantine
equation has a non-negative integer solution.

Proving the relationship in the opposite direction, namely that if Hilbert's
tenth problem can be solved then will be Turing halting problem, is much harder
and requires the so-called Davis-Putnam-Robinson-Matiyasevich (DPRM)
Theorem~\cite{hilbert10}:
\begin{quotation}
\em
Every recursively enumerable (r.e.) set~\footnote{A set $\cal S$ is recursively 
enumerable if it is the range
of an unary recursive function $f: {\mathbb N} \to {\cal S}$.  In other words,
a set is recursively enumerable if there exists a Turing program which semi-computes
it--that is, when provided with an input, the program returns 1 if that input is an
element of the set, and diverges otherwise.} 
of n-tuple of non-negative
integers has a Diophantine representation.  That is, for every such r.e.  set there is a
unique family of Diophantine equations $D(a_1,\ldots,a_n;x_1,\ldots,x_m) = 0$,
each of which has $n$ non-negative integral parameters $(a_1,\ldots,a_n)$
and some $m$ variables $(x_1,\ldots,x_m)$, in such a way that a particular
$n$-tuple $(a_1,\ldots,a_n)$ belongs to the
set if and only if the Diophantine equation corresponding to the same $n$
parameters has some integer solutions~\footnote{When the elements of a set is
not r.e. the DPRM Theorem is not directly
applicable; but in some special cases the Theorem can still be very useful.
One such interesting example is the set whose elements are the positions $n$th of
all the bits of Chaitin's $\Omega$~\cite{Chaitin:1987} which have value 0
(in some fixed programming language).  We refer the readers to~\cite{ordkieu-omega}
for further exploitation of the DPRM Theorem
in representing the bits of $\Omega$ by some properties (be it the parity or the
finitude) of the number of solutions of some Diophantine equations.}.
\end{quotation}
Let us number all Turing machines (that
is, programs in some fixed  programming language) uniquely in some
lexicographical order, say.  The set of all non-negative integer numbers
corresponding to all Turing machines that will halt when started from the blank
tape is clearly a r.e. set.   Let us call this set the halting set, and thanks
to the DPRM Theorem above we know that corresponding to this set there is a
family of one-parameter Diophantine equations.  If Hilbert's tenth problem were
recursively soluble, that is, were there a recursive method to decide if any
given Diophantine equation has any solution then we could have recursively
decided if any Turing machine would halt when started from the blank tape.  We
just need to find the number representing that Turing machine and then decide if
the relevant Diophantine equation having the parameter corresponding to this
number has any solution or not.  It has a solution if and only if the Turing
machine halts.

But that would have contradicted the Cantor's diagonal arguments for the Turing
halting problem!  Thus, one comes to the conclusion that there is no single
recursive method for deciding Hilbert's tenth problem.   For the existence or
lack of solutions of different Diophantine equations one may need different
(recursive) methods anew each time.

Now, having reached this conclusion, we wish to point out that logically there is
nothing wrong if there exist non-recursive or non-deterministic or probabilistic
methods for deciding Hilbert's tenth.

\subsection{\label{jurisdiction}The quantum algorithm in context}
In claiming that our quantum algorithm can somehow compute the noncomputable, we
also need to consider it in the context of the no-go arguments above.
Those arguments, indeed, cannot be applicable here because of several reasons.
First of all, the proof for the working of our algorithm is not quite
constructive, implying its non-recursiveness in some sense.
The mathematical proof for the criterion for
ground-state identification, as can be seen from the Section~\ref{sec:identification},
is highly non-constructive as we have had to employ  at several places the methods
of proof by contradiction and also of
(non-constructive) analysis for continuous functions.

Secondly, and more explicitly, our algorithm is outside the
jurisdiction of those no-go arguments because of its probabilistic in nature.
We have argued elsewhere~\cite{OrdKieu-prob} against the common misunderstanding
that probabilistic computation is equivalent in terms of computability to Turing
recursive computation.  They are not!  We have pointed out that if a non-recursively
biased coin is used as an oracle for a computation, the computation carries
more computability than Turing computation in general.  Here, we shall show 
explicitly in the below how
the probabilistic nature of our algorithm can avoid the Cantor's diagonal
no-go arguments presented in Subsection~\ref{Turing}~\footnote{The seed of the
following arguments was originated from a group discussion with
Enrico Deotto, Ed Farhi, Jeff Goldstone and Sam Gutmann in 2002.  It
goes without saying that if there are mistakes in the interpretation
herein, they are solely mine.}.

Because our algorithm is probabilistic (see Subsection~\ref{subsec:probabilistic}),
we can only obtain an answer with certain probability to be the correct answer.
This probability can be, with more and more work done, made arbitrarily closed, but
never equal, to one as shown in~(\ref{L}).  Thus, instead of the halting
function~(\ref{Cantor}) our algorithm can only yield a probabilistic halting
function $ph$, similar but not quite the same as it must have {\em three} arguments
instead of two,
\begin{eqnarray}
ph(p,i,\delta)&=&\left\{
\begin{tabular}{l}
0 \mbox{ if $p$ halts on input $i$, with maximun error-probability $\delta$};\\
1 \mbox{ if does not halt, with maximun error-probability $\delta$}.
\end{tabular}
\right.
\label{pCantor}
\end{eqnarray}
In order to follow the Cantor's arguments as closely as possible, we need to
restrict to
\begin{eqnarray}
\delta = 2^{-J},
\end{eqnarray}
for some integer $J$.
Following the flow of Cantor's arguments,
one can then construct a program $pTuring(n,\delta)$ having arguments
$n$ and $\delta$ in such a
way that it calls the function $ph(n,n,\delta)$ as a subroutine and then halts
if and only if $ph(n,n,\delta) = 1$.  In some made-up language:
\begin{eqnarray}
\begin{tabular}{ll}
\tt
 &Program $pTuring$\\
 &input ($n$,$\delta$)\\
10& call $ph(n,n,\delta)$\\
 &if $ph(n,n,\delta) = 0$ goto 10\\
 &stop\\
 &end
\end{tabular}
\end{eqnarray}
Similarly, let $t_p$ be the G\"odel encoded integer for $pTuring$ (barring the
case $pTuring$ does not have an integer encoding, which is
quite possible for a quantum algorithm~\cite{kieu-intjtheo}).
We next apply
the probabilistic halting function $ph$ to $t_p$ and $(n,\delta)$, which
is the total input for $pTuring$, and with some $\delta'$, to obtain:
\begin{eqnarray}
ph(t_p,\tilde{n},\delta') = 0 &&
\mbox{\bf iff } pTuring \mbox{ halts on $\tilde n$, with maximum
error-probability $\delta'$},\label{T2}
\end{eqnarray}
where $\tilde{n}$ encodes $(n,\delta)$ uniquely, that is, $\tilde n = p_1^np_2^J$,
where $p_1$ and $p_2$ are two different prime numbers.
No matter what we choose for $n$ and $\delta'$ and $J$, we
{\em cannot diagonalise}
the above, because $\tilde n\not= n$, unlike previously~(\ref{T1}).  
Thus we never run into mathematical contradiction here.

The above arguments apparently have nothing to do with being quantum mechanical,
but everything to do with being probabilistic.  Quantum mechanics, however, has
given us an inspiration to realise a probabilistic algorithm capable of deciding 
Hilbert's tenth problem with a single, universal procedure for any input of 
Diophantine equations.

\section{\label{sec:end}Concluding remarks}
In this paper we have provided the analytical results and arguments for 
the working of our
quantum algorithm for Hilbert's tenth problem.  Numerical simulations for
some simple Diophantine equations have been preliminarily reported
in~\cite{kieu-spie}
and will be available fully elsewhere~\cite{kieu-numer}, where we shall
explain how to cope with the required dimensionally-unbounded Hilbert space
and also argue that while the algorithm has been inspired by (quantum
mechanical)
physical processes it may be simulated on Turing computers, despite the proof
that Hilbert's tenth problem is recursively noncomputable.

\begin{acknowledgments}
I am indebted to Alan Head, Peter Hannaford, Toby Ord and Andrew Rawlinson for 
discussions
and continuing support.  I would also like to acknowledge helpful discussions
with Enrico Deotto, Ed Farhi, Jeff Goldstone and Sam Gutmann during a visit at
MIT in 2002, and
with Peter Drummond and Peter Deuar; 
these discussions have helped sharpen the issues studied in this paper. 
\end{acknowledgments}

\bibliography{adiabatic,ait}

\end{document}